\documentclass[usenatbib]{mn2e}
\usepackage{graphicx,times}
\usepackage{bm}
\usepackage{epsf}
\usepackage{amsmath}
\usepackage{amssymb}
\usepackage{color}

\date{\today,~ $ $Revision: 2$ $}

\def\la{\langle}
\def\ra{\rangle}
\def\n{\noindent}
\def\be{\begin{equation}}
\def\ee{\end{equation}}
\def\ben{\begin{eqnarray}}
\def\een{\end{eqnarray}}
\def\nn{\nonumber}
\def\oh{\hat\Omega}
\def\myC{{\cal C}}

\def\myC{{\cal C}}

\def\2p{{(2\pi)^2}}

\def\be{\begin{equation}}
\def\ee{\end{equation}}
\def\beq{\begin{equation}}
\def\eeq{\end{equation}}
\def\ben{\begin{eqnarray}}
\def\een{\end{eqnarray}}

\def\oh{{\hat\Omega}}
\def\nn{{\nonumber}}

\newcommand{\beqa}{\begin{eqnarray}}
\newcommand{\eeqa}{\end{eqnarray}}

\newcommand{\myf}{{\Theta}}

\def\fw{\theta_b}
\def\nfw{(\theta_b)}
\begin{document}
\onecolumn
\onecolumn
\title[Secondary Anisotropies and Minkowski Functionals]
{Probing CMB Secondary Anisotropies through Minkowski Functionals}
\author[Munshi, Coles \& Heavens ]
{Dipak Munshi$^{1}$, Peter Coles$^{1}$ and Alan Heavens$^{2,3}$\\
${}^1$ School of Physics and Astronomy, Cardiff University, Queen's
Buildings, 5 The Parade, Cardiff, CF24 3AA, UK\\
$^{2}$Imperial Centre for Inference and  Cosmology, Blackett Laboratory, Prince  Consort Road, London SW7 2AZ, UK\\
$^{3}$ Scottish Universities Physics Alliance (SUPA),~ Institute for Astronomy, University of Edinburgh,
Blackford Hill,  Edinburgh EH9 3HJ, UK}
\maketitle
\begin{abstract}
{\em Secondary} contributions to the anisotropy of the Cosmic
Microwave Background (CMB), such as the integrated Sachs-Wolfe (ISW)
effect, the thermal Sunyaev-Zel'dovich effect (tSZ), and the effect
of gravitational lensing, have distinctive non-Gaussian
signatures, and full descriptions therefore require information
beyond that contained in their power spectra. The Minkowski
Functionals (MF) are well-known as tools for quantifying any
departure from Gaussianity and are affected by noise and other
sources of confusion in a different way from the usual methods based
on higher-order moments or polyspectra, thus providing complementary
tools for CMB analysis and cross-validation of results. In this
paper we use the recently introduced skew-spectra associated with
the MFs to probe the topology of CMB maps to probe the secondary
non-Gaussianity as a function of beam-smoothing in order to separate
various contributions. We devise estimators for these spectra in the
presence of a realistic observational masks and present expressions
for their covariance as a function of instrumental noise. Specific
results are derived for the mixed ISW-lensing and tSZ-lensing
bispectra as well as contamination due to point sources for noise
levels that correspond to the Planck ($143$ GHz channel) and EPIC
($150$ GHz channel) experiments. The cumulative signal to noise
ration $S/N$ for one-point generalized skewness-parameters can reach
an order of ${\cal O}(10)$ for Planck and two orders of magnitude
higher for EPIC,  i.e. ${\cal O}(10^3)$. We also find that these
three spectra skew-spectra are correlated, having  correlation
coefficients $r \sim 0.5-1.0$; higher $l$ modes are more strongly
correlated. Though the values of $S/N$ increase with decreasing
noise, the triplets of skew-spectra that determine the MFs become
more correlated; the $S/N$ ratios of lensing-induced skew-spectra
are smaller compared to that of a frequency-cleaned tSZ map.
\end{abstract}

\section{Introduction}
All-sky multi-frequency Cosmic Microwave Background (CMB) missions,
such as the completed WMAP\footnote {http://map.gsfc.nasa.gov/},
ongoing Planck\footnote
{http://www.rssd.esa.int/index.php?project=Planck} and  future
(proposed) Experimental Probe of Inflationary Cosmology (EPIC)
survey \citep{Bock08,Bock09,Bau09} or ESAs Cosmic Origin Explorer (COrE, \cite{Core12})
are major sources of information
about the properties of the primordial density fluctuations that
seeded the process of galaxy formation in the Universe as well as
other key aspects of cosmological theory, including the global
isotropy \citep{Copi07,Hoft09,HanLew09} and topology of the Universe
\citep{Lum03,Rouk04}.

The study of non-Gaussianity in the CMB fluctuations can provide
valuable and detailed information regarding the physics of the early
Universe of the inflationary epoch In the standard slow-roll
paradigm, the scalar field responsible for inflation fluctuates with
a minimal amount of self interaction which ensures that any
non-Gaussianity generated during the inflation through
self-interaction is expected to be small
\citep{Salopek90,Salopek91,Falk93,Gangui94,Acq03,Mal03}; see
\cite{Bartolo06} for a review. Variants of the simple inflationary
model such as multiple scalar fields, features in the inflationary
potential, non-adiabatic fluctuations, non-standard kinetic terms,
warm inflation, or deviations from Bunch-Davies vacuum can however
all lead to higher level of primordial non-Gaussianity
\citep{Chen10}.

However, the detection of departure from Gaussianity in the CMB can
be due to either primary or secondary effects (or both), as well as
the mode-coupling effects of secondaries and gravitational lensing
along the observer's light cone. Secondary anisotropies resulting
from the formation of structure are known to dominate at smaller
angular scales, are highly non-Gaussian in nature
\citep{Cooray01b,CoorayHu,VS02} and are arguably as interesting as
their primary counterpart. One of the prominent contributions to the
secondary non-Gaussianity is due to the mode-coupling of weak
gravitational lensing and sources of secondary contributions such as
the thermal Sunyaev-Zel'dovich effect
\citep{GoldbergSpergel99a,GoldbergSpergel99b, CoorayHu}. Although
weak lensing  of the CMB produces its own characteristic signature
in the angular power spectrum, its detection has proved to be
difficult using the CMB power spectrum alone. Non-Gaussianity
imprinted by lensing into the primordial CMB remains below the
detection level of current experiments, although with Planck the
situation is likely to improve. Nevertheless, cross-correlating CMB
data with external tracers means lensing signals can be probed at
the level of the mixed bispectrum. After the first unsuccessful
attempt to cross-correlate WMAP against SDSS, recent efforts by
\cite{SmZaDo00} have found a clear signal of weak lensing of the
CMB, by cross-correlating WMAP against NVSS. Their work also
underlines the link between three-point statistical estimators and
the estimators for weak lensing effects on CMB. The understanding of
secondaries are not only important in their own right, but also from
the perspective of their impact on estimation of cosmological
parameters \citep{Smidt10}.

The study of non-Gaussianity is usually primarily focused on the
bispectrum, as this saturates the Cram\'er-Rao bound
\citep{Babich,KSH11} and is therefore in a sense optimal, however in
practice it is difficult to probe the entire configuration
dependence in harmonic space contained within the bispectrum using
noisy data \citep{MuHe10}. The cumulant correlators are multi-point
correlators collapsed to encode two-point statistics. These were
introduced in the context of analyzing galaxy clustering by
\cite{Szapudi}, and were later found to be useful for analyzing
projected surveys such as the APM galaxy survey\citep{Mun}. Being
two-point statistics they can be analyzed in multipole space by
defining an associated power-spectrum. Recent studies by
\cite{Cooray3} and \cite{Cooray8} have demonstrated their wider
applicability including, e.g., in 21cm studies. In more recent
studies the skew- and kurt-spectra were found to be useful for
analysing temperature \citep{MuHe10} as well as polarization maps
\citep{Mu11c} from CMB experiments and in weak lensing studies
\citep{Mu11b,Mu11d}.

In addition to studies involving lower order multi-spectra, MFs have
been extensively developed as a statistical tool for non-Gaussianity
in a cosmological setting for both 2-dimensional (projected) and
3-dimensional (redshift) surveys. Analytic results are known certain
properties of the MFs of  a Gaussian random field making them
suitable for identifying non-Gaussianity. Examples of such studies
include CMB data \citep{Schmalzing98,Novikov00,HikageM08,Natoli10},
weak lensing (\cite{Matsubara01,Sato01,Taruya02,MuWaSmCo12}),
large-scale structure
\citep{Gott86,Coles88,Gott89,Melott89,Gott90,Moore92,Gott92,Canavezes98,SSS98,
Schmalzing00,Kerscher01,Hikage02,Park05,Hikage06,Hikage08}, 21cm
\citep{Gleser06}, frequency cleaned Sunyaev-Zel'dovich (SZ) maps
\citep{MuSmJoCo12} and N-body simulations
\citep{Schmalzing00,Kerscher01}. The MFs are spatially-defined
topological statistics and, by definition, contain statistical
information of all orders in the moments. This makes them
complementary to the poly-spectra methods that are defined in
Fourier space. It is also possible that the two approaches will be
sensitive to different aspects of non-Gaussianity and systematic
effects although in the weakly non-Gaussian limit it has been shown
that the MFs  reduce to a weighted probe of the bispectrum
(\cite{Hikage06}).

The skew-spectrum is a weighted statistic that can be tuned to a
particular form of non-Gaussianity, such as that which may arise
either during inflation at an early stage or from structure
formation at a later time. The skew-spectrum retains more
information about the specific form of non-Gaussianity than the
(one-point) skewness parameter alone. This allows not only the
exploration of primary and secondary non-Gaussianity but also the
residuals from galactic foreground and unresolved point sources. The
skew-spectrum is directly related to the lowest order cumulant
correlator and is also known as the two-to-one spectra in the
literature \citep{Cooray01}. In a series of recent publications the
concept of skew-spectra was generalized to analyse the morphological
properties of cosmological data sets or in particular the MFs by
\cite{MSC10,MuWaSmCo12,MuSmJoCo12,PratMun12}. The first of these
three spectra, in the context of secondary-lensing correlation
studies, was introduced by \cite{MuVaCoHe11} and was subsequently
used to analyse data release from WMAP  by \cite{Cala10}.

The primary aim of this paper is to consider the entire set of
generalised skew-spectra resulting from the mode-coupling  of
secondary anisotropies and lensing of the CMB and the contribution
thereof to non-Gaussian morphology of the CMB maps. We will be
considering three different secondary-lensing correlation bispectra.
The secondaries that we consider are the Integrated Sachs-Wolfe
effect (ISW) that dominates at large angular scales \citep{Cooray02}
and the thermal Sunyaev-Zel'dovich (tSZ) effect that dominates at
smaller angular scales \cite{Birk99}. In addition we consider a
foreground, namely the contribution from unresolved point sources.
We will consider two experimental setups, the ongoing Planck
satellite and the the proposed EPIC satellite mission discussed
above.

The layout of the paper is as follows. In \textsection\ref{sec:ani}
we briefly outline the bispectrum corresponding to lensing-secondary
mode-coupling. Next, in \textsection\ref{sec:mf}, we review the
formalism underlying the Minkowski Functionals and in
\textsection\ref{sec:skew_spec} we introduce the generalised
skew-spectra associated with the MFs. In \textsection\ref{sec:estim}
we present the estimators for these spectra and their covariances.
Finally, in \textsection\ref{sec:concl} we discuss our results and
comment on future implementation.

Throughout we will use the  parameters of a WMAP cosmology \citep{Lar11}.

\section{Mode Coupling induced by Lensing - Secondary Cross-correlation and the resulting Bispectrum }
\label{sec:ani}

\begin{figure}
\begin{center}
{\epsfxsize=15 cm \epsfysize=5.5 cm {\epsfbox[27 511 589 714]{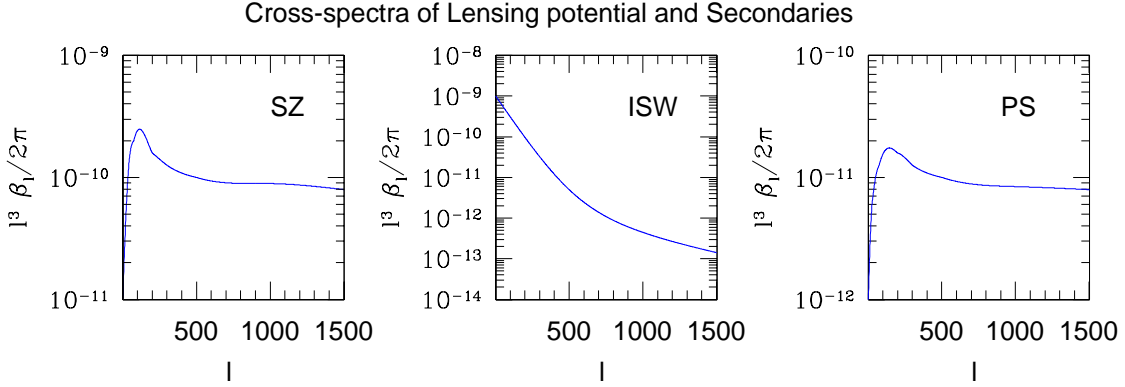}}}
\end{center}
\caption{The cross-spectra $\beta_l$ for various secondaries and lensing are plotted as a function of the harmonics $l$.
From left to right different panels correspond to cross-correlation of lensing potential and SZ, ISW and PS
contribution. The cross-spectra $\beta_l$ is being used in Eq.(\ref{eq:bispec_intro}) for the construction of mixed bispectrum
$B^{\rm PLS}_{l_1l_2l_3}$. It is defined in Eq.(\ref{eq:cross_spec_def}). Various estimators for the skew-spectra that we will use,
associated with the three MFs, will be defined using the mixed bispectra. A background $\Lambda$CDM cosmology
is assumed. The details of these calculations, which relies on halo model prescriptions, can be found in \citep{Cooray01}.}
\label{fig:bls}
\end{figure}

%
The bispectrum of primary anistropies encodes information that can
be used to constrain the inflationary dynamics but, as discussed in
the previous section, the primary contribution to non-Gaussianity is
expected to be negligible in the simplest realisations of the
generic inflationary scenario.

The secondary bispectrum provides valuable information regarding the
low-redshift Universe and constrains structure formation scenarios.
The secondaries can be broadly divided into three different types:
\begin{enumerate}
\item \emph{Gravitational} secondaries, caused by evolution in the
gravitational potential along the observer's past light cone
including the well-known integrated Sachs-Wolfe (ISW) effect
\citep{KS85,MSS90, MFB92,KS94,MSS95,Mo95,BC04} as well as the
Rees-Sciama (RS) effect.

\item \emph{Scattering} secondaries, such as the thermal Sunyaev-Zel'dovich
(tSZ) effect \citep{Birk99}, kinetic Sunyaev Zel'dovich (kSZ) effect
and the Ostriker-Vishniac effect (see e.g. \cite{Ca04}). These effects are caused
by the interaction of the CMB photons with the intervening
free-electron population.

\item \emph{Lensing} secondaries caused by the propagation of photons through large scale structures.

\end{enumerate}

Contributions to secondary bispectra can also arise from terms
involving the cross-correlation of gravitational lensing and the
effects of intervening material, such as the tSZ effect due to
inverse Compton scattering of CMB photons from hot gas in the
intervening clusters. The decay of the peculiar gravitational
potential along the line of sight in $\Lambda$CDM cosmology,
introduced above as the Integrated Sachs-Wolfe or ISW effect, is
correlated to the lensing due to the potential, can also generate an
additional contribution to the secondary bispectrum in a similar
fashion; see e.g. \cite{CooSeth02} for a detailed discussion of
various secondaries in the context of halo model. The contribution
to secondaries due to reionization of the Universe are detailed in
\cite{HKS94}.  Foregrounds, such as unresolved point sources (PS),
can also contribute to the secondary bispectrum through their
cross-correlation with the lensing of CMB.

On a different note, we comment that while the study of secondary
anisotropies is important in their own right, they are also
important in their effect on the calculation of error covariances in
cosmological parameter estimation \citep{Jou10}. Understanding the
detailed statistical properties of secondary anisotropies like those
we discuss here is therefore is of the utmost importance in the era
of precision cosmology.

We will be dealing with the secondary bispectra involving the
lensing of both primary anistropies and other secondaries. Following
\citet{GoldbergSpergel99a}, \citet{GoldbergSpergel99b} and
\citet{CoorayHu} we start by expanding the observed temperature
anisotropy in terms of the primary contribution $\Theta_{\rm
P}(\oh)$, the secondary contribution $\Theta_{\rm S}(\oh)$ and
lensing of the primary $\Theta_{\rm L}(\oh)$: \be \Theta(\oh) =
\Theta_{\rm P} (\oh)+ \Theta_{\rm L} (\oh)+ \Theta_{\rm S}
(\oh)+\cdots. \ee Here $\oh=(\theta,\phi)$ is the angular position
on the surface of the sky. Expanding the respective contribution in
terms of spherical harmonics $Y_{lm}(\oh)$ we can write: \ben
\Theta_{\mathrm {P}}(\oh) \equiv \sum_{lm} (\Theta_{\rm p})_{lm}
Y_{lm} (\oh); ~~~~ \Theta_{\mathrm {L}}(\oh) \equiv \sum_{lm}
[\nabla \psi(\oh) \cdot \nabla \Theta_{\mathrm {P}}(\oh)]_{lm}\;
Y_{lm}(\oh); ~~~~ \Theta_{\mathrm {S}}(\oh) \equiv \sum_{lm}
(\Theta_{\rm S})_{lm} Y_{lm} (\oh). \een Here $\psi(\oh)$ is the
projected lensing potential
\citep{GoldbergSpergel99a,GoldbergSpergel99b}. The secondary
bispectrum for the CMB takes contributions from products of P, L and
S terms with varying order. The bispectrum $B_{l_1l_2l_3}^{\rm PLS}$
are defined as follows (see \cite{Bartolo04} for generic discussion
on the bispectrum and its symmetry properties):
\begin{eqnarray}
B_{l_1l_2l_3}^{\rm PLS} && \equiv \sum_{m_1m_2m_3} \left ( \begin{array}{ c c c }
     l_1 & l_2 & l_3 \\
     m_1 & m_2 & m_3
  \end{array} \right) \int \left \langle \Theta_{\rm P}(\oh_1) \Theta_{\rm L}(\oh_2) \Theta_{\rm S}(\oh_3) \right \rangle
Y^*_{l_1m_1}(\oh_1) Y^*_{l_2m_2}(\oh_2) Y^*_{l_3m_3}(\oh_3) d \oh_1 d \oh_2 d \oh_3; \nonumber \\
&& \equiv \sum_{m_1m_2m_3} \left ( \begin{array}{ c c c }
     l_1 & l_2 & l_3 \\
     m_1 & m_2 & m_3
  \end{array} \right) \langle (\Theta_{\mathrm {\rm P}})_{l_1m_1} (\Theta_{\mathrm {\rm L}})_{l_2m_2}
(\Theta_{\mathrm {\rm S}})_{l_3m_3}  \rangle.
\label{eq:bispec_gen}
\end{eqnarray}
The angular brackets represent {\em ensemble} averages.
The matrices denote $3j$ symbols \citep{Ed68} and the asterisks denote complex conjugation.
It is possible to invert the relation assuming isotropy of the background Universe:
\begin{equation}
\langle (\Theta_{\mathrm {P}})_{l_1m_1} (\Theta_{\mathrm {L}})_{l_2m_2}
(\Theta_{\mathrm {S}})_{l_3m_3}  \rangle = \left ( \begin{array}{ c c c }
     l_1 & l_2 & l_3 \\
     m_1 & m_2 & m_3
  \end{array} \right) B^{\rm PLS}_{l_1l_2l_3}.
\end{equation}
Finally the bispectrum $B_{l_1l_2l_3}^{\rm PLS}$ is expressed in
terms of the un-lensed primary power spectrum ${\cal
C}_l=\la(\Theta_{\rm P})_{lm}(\Theta_{\rm P}^*)_{lm}\ra$ and the
cross-spectra $\beta_l$ (to be defined below) as follows: \ben &&
B_{l_1l_2l_3}^{\rm PLS} = -\left \{ \beta_{l_3} {\cal C}_{l_1} {
l_2(l_2+1) - l_1(l_1+1) - l_3(l_3+1) \over 2 }+ {\rm cyc.perm.}
\right \}
I_{l_1l_2l_3} \equiv {\cal B}_{l_1l_2l_3}I_{l_1l_2l_3}; \label{eq:bispec_intro}\\
&& I_{l_1l_2l_3} \equiv\sqrt {(2l_1 +1)(2l_2+1)(2l_3+1) \over 4\pi  }\left ( \begin{array}{ c c c }
    l_1 & l_2 & l_3 \\
     0 & 0 & 0
  \end{array} \right). 
\label{eq:bispec_intro1} \een (see \citet{GoldbergSpergel99a},
\citet{GoldbergSpergel99b} for a derivation). The reduced bispectrum
above is denoted ${\cal B}_{l_1l_2l_3}$. To simplify the notation
for the rest of this paper, we henceforth drop the superscript $\rm
PLS$ form the bispectrum $B_{l_1l_2l_3}$. The cross-spectrum
$\beta_{l}$ introduced above represents the cross-correlation
between the projected lensing potential $\psi(\oh)$ and the
secondary contribution $\Theta_{\rm S}(\oh)$: \be
\la\psi(\oh)\Theta_{\rm S}(\oh')\ra = {1\over
4\pi}\sum_{l=2}^{l_{\rm max}}(2l+1)\beta_l
P_l(\hat\Omega\cdot\hat\Omega'). \label{eq:cross_spec_def} \ee The
cross-spectra $\beta_l$ take different forms for ISW-lensing,
RS-lensing or SZ-lensing correlation and we assume a zero primordial
non-Gaussianity. The reduced bispectrum ${\cal B}_{l_1l_2l_3}$
defined above using the notation $I_{l_1l_2l_3}$ is useful in
separating the angular dependence from the dependence on power
spectra $C_l$ and $\beta_l$. We will use this to express the
topological properties of the CMB maps. The $\beta_l$ parameters for
lensing secondary correlations are displayed in
Figure-\ref{fig:bls}. The left, middle and right panels in
Figure-\ref{fig:bls} displays SZ-lensing, ISW-lensing and point
source lensing correlations. These results are based on halo model
calculations performed using the halo model \citep{Cooray01}.

The beam $b_l(\theta_b)$ and the noise of a specific experiment are
characterised by the parameters $\sigma_{\rm beam}$ and $\sigma_{\rm
rms}$: \be b_l(\theta_b) = \exp[-l(l+1)\sigma_{\rm beam}^2]; \;\;\;
\sigma_{\rm beam} = {\theta_{b} \over \sqrt{8\ln(2)}}; \;\;\; n_l =
\sigma_{\rm rms}^2\Omega_{\rm pix}; \quad \Omega_{\rm pix} = {4\pi
\over {\rm N}_{\rm pix}}. \label{eq:beam_noise} \ee where
$\sigma_{\rm rms}$ is the rms noise per pixel that depends on the
full width at half maxima or FWHM of the beam $\theta_{b}$. The
number of pixel ${\rm N}_{\rm pix}$ required to cover the sky
determines the size of the pixels $\Omega_{\rm pix}$. To incorporate
the effect of experimental noise and the beam we have to replace
${\cal C}_l \rightarrow {\cal C}_l b_l^2(\theta_b) + n_l$, and the
normalization of the skew-spectra that we will introduce later will
be affected by the experimental beam and noise. The computation of
scatter will also depend on these parameters. We will consider two
different experimental setups: Planck and EPIC. The parameters of
these experiments are tabulated in Table \ref{tab:exp}.

The optimal estimators for lensing-secondary mode-coupling
bispectrum have been recently discussed by \cite{MuVaCoHe11}. The
estimators that we propose here are relevant in the context of
constructing the MFs.

\section{Minkowski functionals}
\label{sec:mf}

Integral geometry provides a natural framework within which to
define the set of morphological descriptors for a random field.
These descriptors are intrinsically defined in the spatial domain
where they take into account all $n$-point correlators up to
arbitrary order. Hadwiger's characterization theorem shows that a
linear combination of these $d+1$ functionals will provide a
complete morphological description of the morphology of
$d$-dimensional objects; see \cite{Hadwiger59} for a formal
treatment. These functionals are more commonly referred to as the
Minkowski functionals. The Minkowski Functionals are usually
calculated using volume-weighted curvature integrals for which the
analytical results for a Gaussian random field are known
\citep{Adler81,Tomita86,Gott90}. More recently the analytical values
for weakly non-Gaussian fields have been calculated as a function of
skewness parameters by using a perturbative approach based on the
Edgeworth expansion
\citep{mat94,Matsubara95,MatsubaraY96,Matsubara02,Hikage06}). This
approach allows us to use the MFs as a test of non-Gaussianity in
the weakly perturbed regime as constrained by observation and
predicted by models for inflation.
\begin{figure}
\begin{center}
{\epsfxsize=9 cm \epsfysize=5. cm {\epsfbox[27 416 587 714]{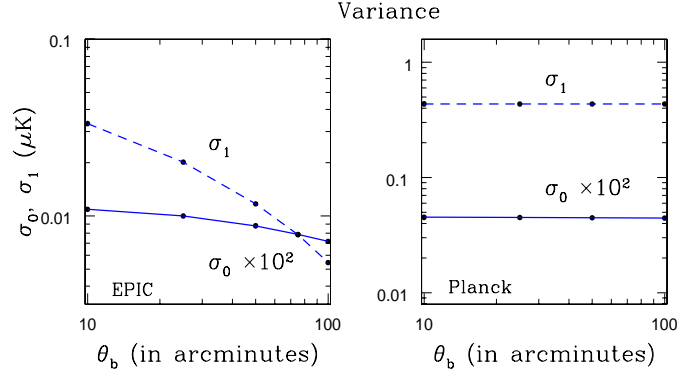}}}
\end{center}
\caption{The variances $\sigma_0(\theta_{b})$ and $\sigma_1(\theta_{b})$, defined in Eq.(\ref{eq:sigma0}), are plotted as a function of the FWHM $\theta_{b}$.
The left panel corresponds to an experimental setup such as EPIC and the right panel corresponds to Planck-type
experiment.See Table-\ref{tab:exp} for detail specifications regarding the noise level and beam.}
\label{fig:sigma}
\end{figure}

In 2 dimensions the MFs $V_0(\nu)$, $V_1(\nu)$ and $V_2(\nu)$
correspond respectively to the area of a set $\Sigma$, length of the
perimeter of the set and the integrated curvature along its
boundary. The MF $V_2(\nu)$ can be related to the well-known genus
$g$ and the Euler characteristic $\chi$: \be V_0(\nu) =
\int_{\Sigma} da; \quad V_1(\nu) = {1 \over 4}\int_{\partial\Sigma}
dl; \quad V_2(\nu) = {1 \over 2\pi}\int_{\partial \Sigma}\kappa dl.
\ee Here $dl$ and $da$ represent the length and surface element
respectively.
In our analysis we consider a smoothed random field $\myf(\oh)$ with
mean $\la \myf(\oh)\ra=0$ and variance $\sigma_0^2(\theta_b) = \la
\myf^2(\oh) \ra$. For a generic 2-dimensional weakly non-Gaussian
random field $\myf$ on the surface of the sky,  the spherical
harmonic decomposition using $Y_{lm}(\oh)$ as basis functions
$\myf(\oh) = \sum_{lm} \myf_{lm} Y_{lm}(\oh)$ can be used to define
the power spectrum $\myC_l$ which is sufficient to characterize an
isotropic Gaussian field  $\la \myf_{lm} \myf^*_{l'm'} \ra = \myC_l
\delta_{ll'}\delta_{mm'}$.

We will be studying the MFs defined over the surface of the
celestial sphere but equivalent results can be obtained in 3D using
a Fourier decomposition \citep{PratMun12}. The MFs for a 3D random
Gaussian field are well known and are given by Tomita's formula
\citep{Tomita86}.

For a non-Gaussian field the higher order statistics such as bi- or
tri-spectrum can describe the resulting mode-mode coupling.
Alternatively topological measures such as the MFs (including the
Euler characteristic or genus) can be employed to quantify
deviations from Gaussianity. Indeed it can be shown that the
information content in both descriptions is equivalent in that, at
leading order, the MFs can be constructed completely from the
knowledge of the bispectrum alone.

The notations and analytical results in this section are being kept
generic however they will be specialized to the case of CMB sky in
subsequent discussions.

The MFs denoted as $V_k(\nu)$ for a threshold $\nu= \myf/\sigma_0$;
where $\sigma^2_0(\theta_b)=\la \myf^2 \ra$ are perturbatively
expressed as: \ben && V_k(\nu) = {1 \over (2\pi)^{(k+1)/2}}
{\omega_2 \over \omega_{2-k}\omega_k} \exp \left ( -{\nu^2 \over
2}\right ) \left ( \sigma_1 \over \sqrt 2 \sigma_0 \right )^k
\left [ V_k^{(2)}(\nu)\sigma_0(\theta_b) + V_k^{(3)}(\nu)\sigma_0^2(\theta_b) + V_k^{(4)}(\nu)\sigma_0^3 (\theta_b)+ \cdots \right ] \label{eq:v_k1}\\
&& V_k^{(2)}(\nu) = \left [ \left \{  {1 \over 6} S^{(0)}(\theta_{b}) H_{k+2}(\nu) + {k \over 3} S^{(1)}(\theta_{b}) H_k(\nu) +
{k(k-1) \over 6} S^{(2)}(\theta_{b}) H_{k-2}(\nu)\right \} \right ] \label{eq:v_k2} \\
&& \sigma_j^2(\theta_b) = {1 \over 4\pi }\sum_l (2l+1) [l(l+1)]^j \myC_l b_l^2(\theta_{b})
\label{eq:sigma0}
\een

The constant $\omega_k$ introduced above is the volume of the unit
sphere in $k$ dimensions. $w_k = {\pi^{k/2}/ \Gamma(k/2+1)}$ in
2-dimension we will only need $\omega_0=1$, $\omega_1=2$ and
$\omega_2 =\pi$. Here $\Gamma$ is the the gamma function. The
lowest-order Hermite polynomials $H_k(\nu)$ are listed below.  \ben
&& H_{-1}(\nu) = \sqrt{\pi \over 2} \exp\left ({\nu^2 \over 2}\right ) {\rm erfc} \left (\nu \over \sqrt 2 \right ); \quad H_0(\nu) = 1, \quad H_1(\nu) = \nu, \nn \\
&& H_2(\nu)=\nu^2 -1, \quad H_3(\nu)=\nu^3 - 3\nu, \quad \quad H_4(\nu) = \nu^4 - 6\nu^2 + 3, \nn \\
&& H_n(\nu) = (-1)^n \exp \left ({ \nu^2 \over 2 } \right ) {d \over
d\nu^n} \exp \left (-{\nu^2 \over 2 }\right ). \een The expression
consists of two distinct contributions. The part that does not
depend on the three different skewness parameters
$S^{(0)}(\theta_{b}), S^{(1)}(\theta_{b}), S^{(2)}(\theta_{b})$ and
signifies the MFs for a Gaussian random field. The other
contribution $\delta V_k(\nu)$ represents the departure from the
Gaussian statistics and depends on the generalized skewness
parameters defined in Eq.(\ref{skew1}) - Eq.(\ref{skew3}).
\begin{figure}
\begin{center}perturbatively
{\epsfxsize=15 cm \epsfysize=5 cm {\epsfbox[27 511 589 714]{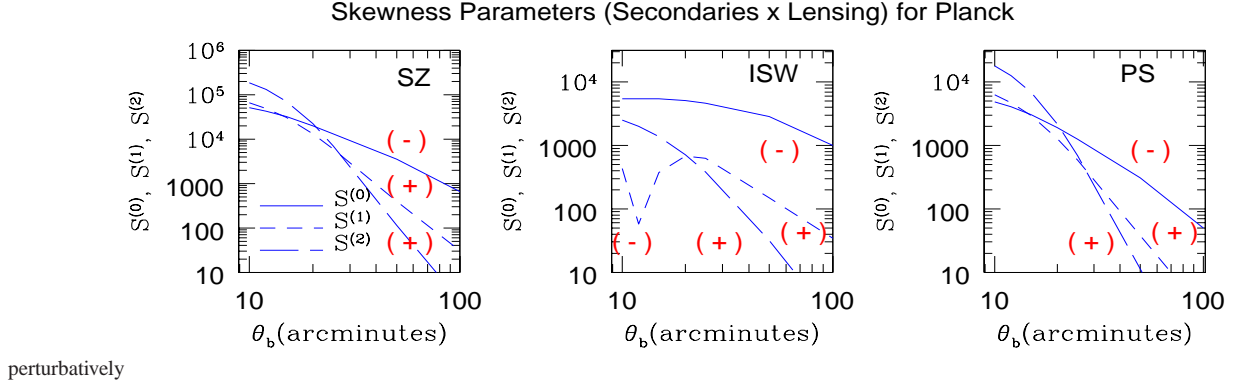}}}
\end{center}
\caption{Same as previous plot but for Planck type noise. The normalisation of the skew spectra is depends on the
noise level. The shape of the bispectrum is independent of the noise but depends on the experimental beam.
Notice that $S^{(1)}(\theta_{b})$ changes sign near $\theta_{b}\sim10'$ as before. $S^{(0)}(\theta_b)$ is negative
for all three effects. The other parameters remain positive for the entire range of FWHM probed. }
\label{fig:skewness_planck}
\end{figure}
\begin{figure}
\begin{center}
{\epsfxsize=15 cm \epsfysize=5 cm {\epsfbox[27 511 589 714]{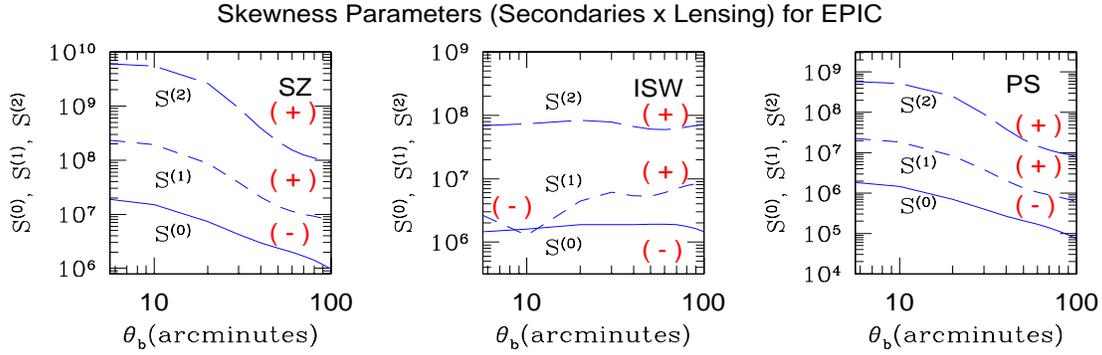}}}
\end{center}
\caption{The absolute magnitudes of the one-point skewness parameters $S^{(0)}(\theta_{b})$, $S^{(1)}(\theta_{b})$ and
$S^{(2)}(\theta_{b})$ (solid, short-dashed and long dashed lines respectively), are plotted as a
 function of smoothing scale $\theta_{b}$. These parameters are defined in Eq.(\ref{skewness_real_space}).
Various panels correspond to contributions from different types of secondary non-Gaussianity
that represent the cross-correlation of lensing and Sunyaev-Zel'dovich effect (left-panel),
cross-correlation of ISW and lensing (middle-panel) and cross-correlation of point sources and lensing (right-panel).
The parameter $S^{(0)}(\theta_b)$ is negative for the entire range of FWHM or $\theta_b$ considered, for all three bispectra we have studied.
The parameter $S^{(1)}(\theta_b)$ shows a transition from being negative to positive for the case of ISW bispetcrum.
All other skewness parameters stays positive as indicated.
An experimental set up which is same as EPIC was considered. See Table-\ref{tab:exp}
for detailed specifications regarding level of noise and beam. The bispectrum used in our calculation is given in Eq.(\ref{eq:bispec_intro}) and
the cross-spectra are plotted in Figure-\ref{fig:bls}.}
\label{fig:skewness_epic}
\end{figure}
Various second-order moments $\sigma_j(\theta_{b})$ defined in
Eq.(\ref{eq:sigma0}) appear in Eq.(\ref{eq:v_k1}) and
Eq.(\ref{eq:v_k2}) can be expressed in terms of the power spectra
$\myC_l$ and the observational beam $b_l(\theta_b)$, assumed
Gaussian with a full width at half maximum $\theta_{b}$; see
Eq.(\ref{eq:beam_noise}) for definition. The moment
$\sigma_0(\theta_b)$ is a special case which relates to the variance
of the field. The quantities $\sigma_1(\theta_b)$,
$\sigma_2(\theta_b)$ are natural generalization of this variance,
putting greater weight on higher-order harmonics; the variances that
appear most frequently henceforth are $\sigma_0^2(\theta_b) = \la
\myf^2 \ra $ and $\sigma_1^2(\theta_b) = \la (\nabla \myf)^2 \ra$.

Real space expressions for the triplets of skewness
$S^{(i)}(\theta_b)$ are given below. These are natural
generalizations of the ordinary skewness $S^{(0)}(\theta_b)$ that is
used in many cosmological studies. They are all cubic statistics but
are constructed from different cubic combinations. \ben &&
{S}^{(0)}(\theta_b) \equiv {S^{(\myf^3)}(\theta_b) \over
\sigma_0^4(\theta_b)} = {\la \myf^3 \ra \over \sigma_0^4(\theta_b)};
\quad
{S}^{(1)}(\theta_b) \equiv-{3 \over 4}{S^{(\myf^2\nabla^2 \myf)}\nfw \over \sigma_0^2(\theta_b)\sigma_1^2(\theta_b)} = -{3 \over 4}{\la \myf^2 \nabla^2 \myf \ra \over \sigma_0^2\nfw \sigma_1^2\nfw}; \quad \nn\\
&& S^{(2)}(\theta_b) \equiv -3 {S^{(\nabla \myf \cdot \nabla \myf
\nabla^2 \myf)}(\theta_b)\over \sigma_1^4(\theta_b)} = -{3}{\la
(\nabla \myf).(\nabla \myf) (\nabla^2\myf) \ra  \over
\sigma_1^4(\theta_b)}. \label{skewness_real_space} \een \n Notice
that a knowledge of the $S^{(i)}(\theta_{b})$ parameters completely
specifies the $V_k^{(2)}(\nu)$ parameters which characterize the
lowest order departure from Gaussianity. The expressions in the
harmonic domain are more useful in the context of CMB studies where
these skewness parameters can be recovered from a masked sky using
analytical tools that are commonly used for power spectrum analysis.
The skewness parameter $S^{(1)}(\theta_{b})$ is constructed from the
product field $[\myf^2]$ and $[\nabla^2 \myf]$, whereas skewness
parameter $S^{(2)}(\theta_{b})$ relies on the combination of
$[\nabla \myf\cdot\nabla \myf]$ and $[\nabla^2 \myf]$. By
construction, the skewness parameter $S^{(2)}(\theta_{b})$ has the
highest weight for high $l$ modes and   $S^{(0)}_{b}$ has the lowest
weight from high $l$ modes.

The expressions in terms of the bispectrum $B_{l_1l_2l_3}$ (see
Eq.(\ref{eq:bispec_gen}) for definition) take the following form:
\ben
&& S^{(\myf^3)}(\theta_b) = {1 \over 4 \pi} \sum_{l_i=2}^{l_{max}} B_{l_1l_2l_3}I_{l_1l_2l_3}b_{l_1}(\theta_b)b_{l_2}(\theta_b)b_{l_3}(\theta_b), \label{skew1} \\
&& S^{(\myf^2\nabla^2\myf)}(\theta_b) = -{1 \over 12 \pi} \sum_{l_i=2}^{l_{max}}{} \Big [{l_1(l_1+1)+ l_2(l_2+1)+ l_3(l_3+1)} \Big ] B_{l_1l_2l_3}I_{l_1l_2l_3}b_{l_1}(\theta_b)b_{l_2}(\theta_b)b_{l_3}(\theta_b),  \label{skew2} \\
&& S^{(\nabla \myf\cdot\nabla \myf \nabla^2 \myf)}(\theta_b) = {1 \over 4 \pi} \sum_{l_i=2}^{l_{max}}
{}\Big [ [l_1(l_1+1)+l_2(l_2+1) - l_3(l_3+1)]l_3(l_3+1) + {\rm cyc.perm.} \Big ]
 B_{l_1l_2l_3}I_{l_1l_2l_3}b_{l_1}(\theta_b)b_{l_2}(\theta_b)b_{l_3}(\theta_b).
\label{skew3} \een The dependence of the skew spectra
$S^{(i)}(\theta_b)$ and the beam $b_l(\theta_b)$ on the smoothing
angular scale $\theta_{b}$ is being suppressed for brevity.

The angular bispectrum $B_{l_1l_2l_3}$ contains all the information
at the level of the three-point angular correlation function. These
results are generic and do not assume any specific form of
non-Gaussianity. However, to arrive at specific expressions we will
ignore the primordial non-Gaussianity, known to be sub-dominant, and
concentrate on secondary non-Gaussianity. There is a family of
one-point statistics, namely the well-known skewness parameters
$S^{(i)}(\theta_{b})$ introduced above, or pseudo-collapsed
three-point function \citep{Hinsaw95}, as well as the equilateral
configuration statistics \citep{FMG98} which can all be expressed as
linear combinations of the bispectrum terms. The generalized
skewness parameters introduced above  are also all linear
combinations of the bispectrum weights but with varying weights.
Using one-point statistics has the advantage of higher signal to
noise but the price we pay is in terms of reduced power to
discriminate individual contributions.

The series expansion for the MFs can be extended beyond the leading
order at the level of the bispectrum to the next-to-leading order
which involves the trispectrum of the temperature field. The lensing
induced trispectrum of the CMB will constitute the main
next-to-leading order contribution. It is also important to realize
that measurements of skewness parameters $S^{(0)}(\theta_{b})$,
$S^{(1)}(\theta_{b})$ and $S^{(2)}(\theta_{b})$ will not be
independent but correlated with one another; the level of
correlation depends on the noise and beam profile.

In Figure \ref{fig:sigma} we have plotted the variance parameters
$\sigma^2_0(\theta_{b})$ and $\sigma^2_1(\theta_{b})$ for various
smoothing beams (assumed Gaussian). The four different FWHM that are
considered are $\theta_b = 10', 25', 50'$ and $100'$ respectively.
The parameter values only depend on the underlying CMB power spectra
and the beam as well as the noise. They are used as a normalization
parameters while constructing the MFs from the generalized skewness
parameters. Two different beam and noise levels are considered EPIC
(left-panel) and Planck (right-panel). The one-point generalised
skewness parameters are depicted in Figure-\ref{fig:skewness_planck}
for Planck and Figure-\ref{fig:skewness_epic} for EPIC. The
background cosmology is that of $\Lambda$CDM.

In Figure \ref{fig:skewness_
epic} the one-point skewness parameters $S^{(0)}(\theta_{b})$,
$S^{(1)}(\theta_{b})$ and $S^{(2)}(\theta_{b})$ (solid, short-dashed
and long dashed lines respectively), are plotted as a function of
smoothing scale $\theta_{b}$. These parameters are defined in
Eq.(\ref{skewness_real_space}). The panels correspond to
contributions from different types of secondary non-Gaussianity:
cross-correlation of lensing and Sunyaev-Zel'dovich effect
(left-panel), cross-correlation of ISW and lensing (middle-panel);
cross-correlation of point source and lensing (right-panel). An
experimental set up which is same as EPIC was considered. See Table
\ref{tab:exp} for detail specifications regarding level of noise and
beam. The bispectrum used in our calculation is given in
Eq.(\ref{eq:bispec_intro}) and the cross-spectra is plotted in
Figure \ref{fig:skewness_planck} we plot the one-point skewness
parameters for Planck.

\section{The triplets of Skew-Spectra and Lowest Order Corrections to Gaussian MFs}
\label{sec:skew_spec}
\begin{figure}
\begin{center}
{\epsfxsize=15 cm \epsfysize=4.8 cm {\epsfbox[27 511 589 714]{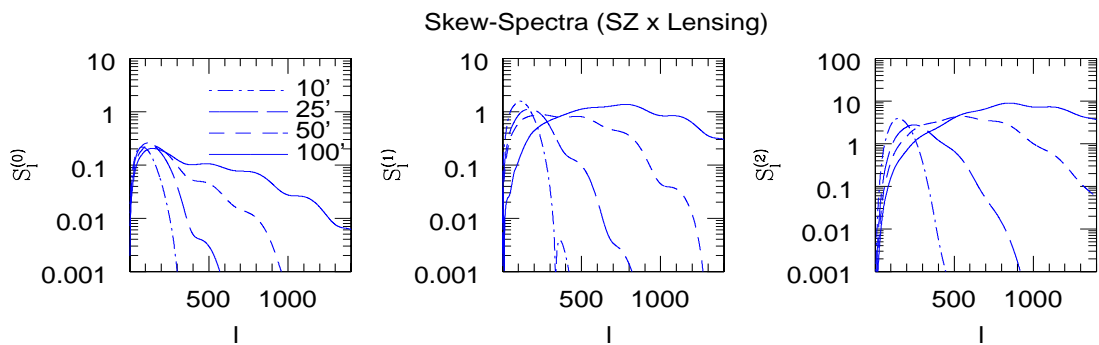}}}
\end{center}
\caption{The three skew-spectra corresponding to the three MFs
$S_l^{(0)}(\theta_{b})$ (left-panel) , $S_l^{(1)}(\theta_{b})$
(middel-panel) and  $S_l^{(2)}(\theta_{b})$ (right-panel) are
plotted for the mixed secondary bispectrum of SZ$\times$Lensing as a
function of the harmonics $l$. A background $\Lambda$CDM cosmology
is assumed.The skew-spectra are defined in Eq(\ref{skew1}),
Eq(\ref{skew2}) and Eq(\ref{skew3}) respectively. All other sources
of non-Gaussianity are ignored. Four different Gaussian-beams are
considered. The curves from left to right correspond to $\theta_{b}
= 10', 25', 50', 100'$ in each panel. The normalization of the skew
spectra is somewhat arbitrary and do not affect the signal to noise
ratios. We have ignored the presence of noise, as defined in
Eq.(\ref{eq:sigma0}), in our calculation of $\sigma_0(\theta_{b})$
and $\sigma_1(\theta_{b})$ respectively. Notice that for a given
$\theta_{b}$ higher order skew-spectra peak at a higher $l$. The
signal to noise of the skew-spectra associated with SZ effect are
plotted in Figure \ref{fig:s2n_epic} for EPIC and Figure
\ref{fig:s2n_planck} for Planck noise level respectively. The
skew-spectra are not uncorrelated. The cross-correlation among
various skew-spectra are displayed in Figure \ref{fig:cross_epic}
for EPIC and Figure \ref{fig:cross_planck} for Planck. }
\label{fig:sim_sz}
\end{figure}
\begin{figure}
\begin{center}
{\epsfxsize=15 cm \epsfysize=4.8 cm {\epsfbox[27 511 589 714]{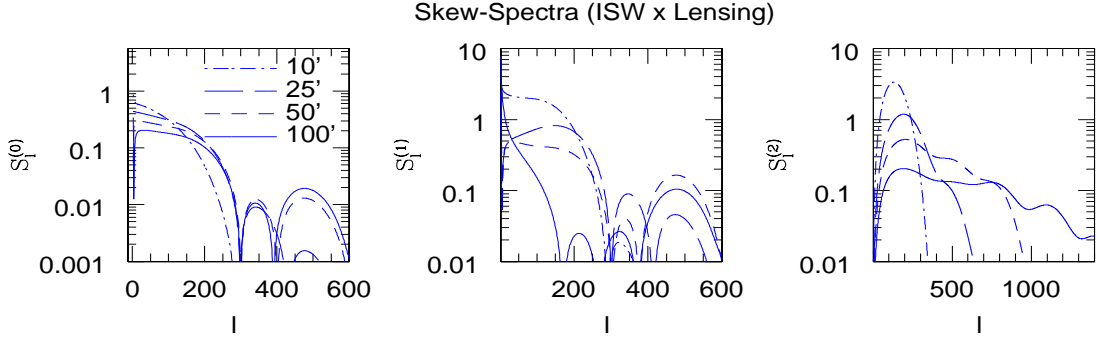}}}
\end{center}
\caption{The skew-spectra $S_l^{(0)}(\theta_{b})$, $S_l^{(1)}(\theta_{b})$ and $S_l^{(2)}(\theta_{b})$ for ISW$\times$Lensing
is plotted as a function of the harmonics $l$. Notice that skew-spectra corresponding to ISW are dominant at smaller $l$s
while the ones corresponding to SZ dominate at larger $l$ values. The skew-spectra, being integrated measures,
depend on the entire harmonic range of the bispectra. The shape of the skew-spectra can play an important role in
separating individual contributions of secondary non-Gaussianity.}
\label{fig:sim_isw}
\end{figure}
The skew-spectrum has been studied previously in various
cosmological contexts \citep{Cooray01}, e.g. to estimate the
bispectrum resulting from lensing-SZ correlation. The skew-spectra
are cubic statistics constructed by cross-correlating two different
fields. One of the fields used is a composite field (map) typically
a  product of two maps either in its original form or constructed by
means of applying relevant differential operators. Example of such
derived maps that we will consider are $[\Theta^2(\oh)],
[\nabla\Theta(\oh)\cdot\nabla\Theta(\oh)]$ and
$[\nabla^2\Theta(\oh)]$. The skew-spectra resulting from
cross-correlating these maps are known as the {\em generalised} skew
spectra and are related to the three {\em generalised} skewness
parameters introduced above. At the lowest order, the MFs themselves
can be constructed using these generalized skewness parameters and
contain equivalent information.

The detection of each individual mode of the primary or secondary
bispectrum is still considered challenging. This is primarily due to
the low signal-to-noise associated with each individual modes. All
available information is therefore typically compressed into a
single number - the skewness. This drastic data compression leads to
a significant degradation of the power of the statistic to
discriminate between models.

The first of the skew-spectra that we will study is the one
introduced by \citep{Cooray01} and later generalized \cite{MuHe10}.
It is related to sometimes known as the two-to-one power spectrum
and is constructed by cross-correlating the squared map
$[\myf^2(\oh)]$ with the original map $\myf(\oh)$. The second
skew-spectrum is constructed by cross-correlating the squared map
$[\myf^2(\oh)]$ with $[\nabla^2\myf(\oh)]$. Analogously the third
skew-spectrum represents the cross-spectra that can be constructed
using $[\nabla \myf(\oh)\cdot\nabla \myf(\oh)]$ and $[\nabla^2
\myf(\oh)]$ maps. \ben && S_l^{(0)}(\theta_b) \equiv {1 \over 12 \pi
\sigma_0^4(\theta_b)}S_l^{(\myf^2,\myf)}(\theta_b) \equiv {1 \over
12 \pi \sigma_0^4(\theta_b)}{1 \over 2l+1}\sum_m {\rm
Real}([\myf]_{lm}[\myf^2]^*_{lm})  ={1 \over 12 \pi
\sigma_0^4(\theta_b)} \sum_{l_1l_2} B_{ll_1l_2}J_{ll_1l_2}
b_{l}(\theta_b)b_{l_1}(\theta_b)b_{l_2}(\theta_b),  \label{sl1} \\
&& S_l^{(1)}(\theta_b) \equiv {1 \over 16 \pi \sigma_0^2(\theta_b)\sigma_1^2(\theta)}S_l^{(\myf^2,\nabla^2 \myf)}(\theta_b)
\equiv {1 \over 16 \pi \sigma_0^2(\theta_b)\sigma_1^2(\theta_b)}{1 \over 2l+1}\sum_m
{\rm Real}([\nabla^2 \myf]_{lm}[\myf^2]^*_{lm}) \nn \\
&&\quad\quad ={1 \over 16 \pi \sigma_0^2(\theta_b)\sigma_1^2(\theta_b)} \sum_{l_i} \Big [{l(l+1)+ l_1(l_1+1)+ l_2(l_2+1)} \Big ] B_{ll_1l_2}J_{ll_1l_2}
b_{l_1}(\theta_b)b_{l_2}(\theta_b)b_{l_3}(\theta_b), \label{sl2} \\
&&
S_l^{(2)}(\theta_b) \equiv {1 \over 8 \pi \sigma_1^4(\theta_b)}S_l^{(\nabla \myf\cdot\nabla \myf, \nabla^2\myf)}(\theta_b) \equiv
{1 \over 8 \pi \sigma_1^4(\theta_b)}{1 \over 2l+1}\sum_m
{\rm Real}([\nabla \myf \cdot \nabla \myf]_{lm}[\nabla^2 \myf]^*_{lm}) \nn \\
&&\quad\quad ={1 \over 8 \pi \sigma_1^4(\theta_b)} \sum_{l_i}
{}\Big [ [l(l+1)+l_1(l_1+1) - l_2(l_2+1)]l_2(l_2+1) + {\rm cyc.perm.} \Big ]
 B_{ll_1l_2}J_{ll_1l_2}b_{l}(\theta_b)b_{l_1}(\theta_b)b_{l_2}(\theta_b) \label{sl3},\\
&& J_{l_1l_2l_3} \equiv {I_{l_1l_2l_3} \over 2l_3+1} = \sqrt{(2l_2+1)(2l_3+1) \over (2l_1+1) 4 \pi }\left ( \begin{array}{ c c c }
     l_1 & l_2 & l_3 \\
     0 & 0 & 0
  \end{array} \right),\\
&& S^{(i)}(\theta_b) = \sum_{l}(2l+1)S^{(i)}_l(\theta_b).
\label{eq:S_l} \een Each of these spectra probes the same bispectrum
$B_{ll_1l_2}$ but with different weights. Each triplet of modes
specifies a triangle in the harmonic domain and the skew-spectra sum
over all possible configuration of the bispectrum keeping one of its
sides $l$ fixed.

The extraction of skew-spectra from data is relatively
straightforward. The procedure consists of the construction the
relevant maps in real space either by algebraic or differential
operations and then cross-correlating them in  multipole space.
Issues related to mask and noise will be dealt with in later
sections. We will show that even in the presence of a mask the
computed skew spectra can be inverted to give a unbiased estimate of
all-sky skew-spectra; presence of noise will only affect the
scatter. We have explicitly displayed the experimental beam $b_l$ in
all our expressions.

\begin{figure}
\begin{center}
{\epsfxsize=15 cm \epsfysize=4.8 cm {\epsfbox[27 511 589 714]{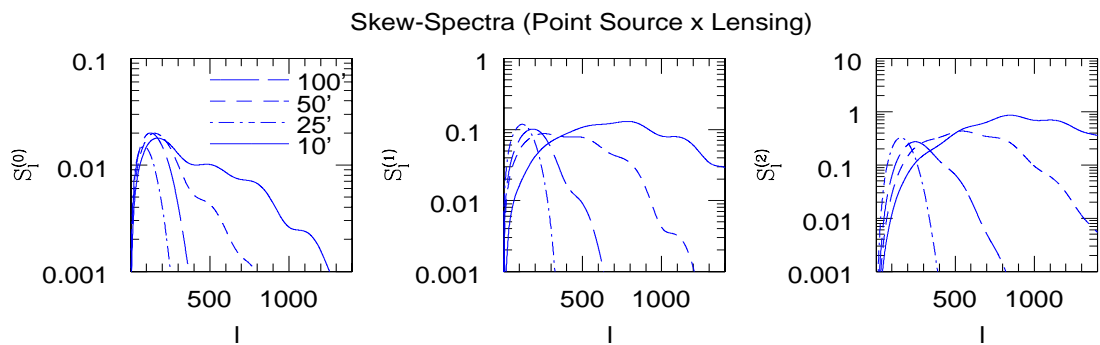}}}
\end{center}
\caption{The skew-spectra $S_l^{(0)}\nfw,\;S_l^{(1)}\nfw$ and $S_l^{(2)}\nfw$ for Point-Source$\times$Lensing
is plotted as a function of the harmonics $l$. The similarity of the underlying bispectrum for unresolved point-sources
and the SZ effect means the shape of resulting skew-spectra are similar.}
\label{fig:sim_ps}
\end{figure}

In Figs:\ref{fig:sim_sz}-\ref{fig:sim_ps}, we have presented the
three different skew-spectra $S_l$ as a function of the harmonics
$l$. The skew-spectra for a generic bispectrum is defined in
Eq.(\ref{sl1}), Eq.(\ref{sl2}) and in Eq.(\ref{sl3}). In Figure
\ref{fig:sim_sz} we present the skew-spectra corresponding to the SZ
effect cross-correlated to lensing. The Figure \ref{fig:sim_isw} we
present the skew-spectra for the ISW effect and Figure
\ref{fig:sim_ps} shows the skew-spectra for unresolved point
sources. The skew spectra are sensitive to the beam $b_l(\theta_b)$
moreover the skew-spectra at a given $l$, i.e $S_l^{(i)}(\theta_b)$
depend on the bispectrum $B_{l_1l_2l_3}$ defined over the entire
range of modes specified by all possible $l$ values that are being
probed. The distinct shape of these individual spectra can be used
to study the nature of their origin (i.e. primordial or secondary).
Specific models of primary non-Gaussianity such as local or
equilateral too will have distinct shapes for the
$S_l^{(i)}(\theta_b)$ parameters though such contributions will be
subdominant for currently allowed levels of primordial
non-Gaussianity.

It is important to stress that these three skew-spectra do not
contain completely independent information; the errors associated
with them are correlated. We next turn to a detailed derivation of
signal-to-noise level of these estimators and the level of
cross-correlation among these spectra for a given observational
strategy. The derivations are accurate for near all-sky coverage;
for more accurate modeling a computationaly expensive but
conceptually straightforward Monte-Carlo analysis is required.

Using the estimator Eq.(\ref{sl1}) previous studies have focused
towards a detection of lensing-secondary correlation for individual
WMAP frequency channels using raw as well as frequency-cleaned maps
\citep{Cala10}. These studies used the KQ75 mask and were limited by the
WMAP resolution $\rm N_{side}=512$ and an $l_{\rm max}=600$. No
significant evidence for a non-Gaussian signal from the
lensing-secondary correlation was found in any of the individual
bands, 2$\sigma$ and 3$\sigma$ evidence were obtained both for
lensing-ISW and lensing-SZ signals in the foreground cleaned Q-band
maps, respectively. They also found that the point source amplitude
at the bispectrum level to be consistent with previous measurements.
With higher resolution maps available from Planck as well as other
future missions such as EPIC it will be possible not only to achieve
a cross validation using multiple skew spectra, but it should also
be possible to reconstruct the topological properties and compare
them with the ones obtained in the pixel domain.

A great deal of attention has recently been focused on designing
optimal estimators. Indeed this is true that for current generation
of experiments (WMAP) the mere detection of non-Gaussianity remains
a challenging task because of the low ratio of signal to noise.
Optimality of an estimator may not be a crucial issue for high
resolution data from experiments such as Planck, at least for the
detection of secondaries, as very high level of signal to noise is
expected. Attention then will shift to the characterization of
non-Gaussianity using an array of estimators to separate different
components of non-Gaussianity (primordial and secondary) and provide
the level of contamination from foregrounds such as point sources.
The skew-spectra associated with MFs can play a valuable role in
this direction. The main advantage of computing the skew-spectra
being a direct estimator which can be computed using a Pseudo-${\cal
C}_{\ell}$ based approach, and the covariances can be characterized
analytically even in the presence of an arbitrary mask and
non-stationary noise.
\section{Estimator, Sky Coverage and Error Analysis}
\label{sec:estim} The results derived above correspond to a
situation in which an all-sky map is available which is free from
noise. However, in reality often we have to deal with issues that
are related to the presence of a mask and (possibly inhomogeneous)
noise. Partial sky coverage introduces mode-mode coupling in the
harmonic domain in such a way that individual masked harmonics
become linear combinations of all-sky harmonics. The coefficients
for this linear transformation depend on specific choice of mask
through its own harmonic coefficients. We will devise a method that
can be used to correct for this mode-mode coupling based on the
Pseudo-${\cal C}_{\ell}$ (PCL) method devised by \cite{Hiv} for
power spectrum analysis and later developed by \cite{MuVaCoHe11} for
analyzing the skew spectra and the kurt-spectrum \citep{Mu11c}.

Consider two generic fields $A(\oh)$ and $B(\oh)$ and denote their
harmonic decompositions in the presence of a mask $w(\oh)$ as
$\tilde A_{lm}$ and $\tilde B_{lm}$. Notice that the mask is
completely general and our results do not depend on any specific
symmetry requirements such as the azimuthal symmetry. The fields $A$
and $B$ may correspond to any of the fields we have considered
above. In a generic situation $A$ and $B$ will denote composite
fields and  the harmonics $A_{lm}$ and $B_{lm}$ will correspond to
any of the harmonics listed in Eq.(\ref{eq:S_l}) i.e.,
$[\myf^2]_{lm}$, \; $[\nabla \myf\cdot \nabla \myf]_{lm}$ and
$[\nabla^2 \myf]_{lm}$:
%
%
\begin{table*}
\begin{center}
\begin{tabular}{|c |c|c| c| c}
\hline
Mission & $\theta_{b}$  & $\sigma_{\rm pix}$ & $\Omega_{\rm pix}$ & Frequency\\
\hline
Planck & $7.1'$ & $2.2 \times 10^{-6}$ & 0.0349 & 143 (GHz)\\
\hline
EPIC & $5.0'$ & $8.0 \times 10^{-9}$ & 0.002 & 150 (GHz)\\
\hline
\end{tabular}
\caption{Parameters used to compute the skew-spectra and the associated scatter for the two
different experiments, ongoing Planck \citep{Planck08} and EPIC \citep{Bau09}.\label{tab:exp}}
\end{center}
\end{table*}
%
%
\ben
&& \tilde A_{lm} = \int ~d\oh~ Y^*_{lm}(\oh)~ [w(\oh)~ A(\oh)]; \\
%
&& \tilde A_{lm} = \sum_{l_im_i} (-1)^m~I_{ll_1l_2} \left ( \begin{array}{ c c c }
     l_1 & l_2 & l \\
     m_1 & m_2 & -m
  \end{array} \right) w_{l_1m_1} A_{l_2m_2}.
\een \n
Similar expressions holds for $B$. The above expression
relates the masked harmonics denoted by $\tilde A_{lm}$ and $\tilde
B_{lm}$  with their all-sky counterparts $A_{lm}$ and $B_{lm}$
respectively. In their derivation we use the Gaunt integral to
express the overlap integrals involving three spherical harmonics in
terms of the $3j$ symbols \citep{Ed68}. The matrix $I_{l_1l_1l_3}$
encodes the overlap integral defined in Eq.(\ref{eq:bispec_intro1}).
The expressions also depend on the harmonics of the mask $w_{lm}$.
If we now denote the (cross) power spectrum constructed from the
masked harmonics and denote it by $\tilde S_l\nfw$ and its all-sky
counterpart by $S_l\nfw$ we can write: \ben && \tilde S_l^{A,B}\nfw
= {1 \over 2l +1} \sum_{m} \tilde A_{lm} \tilde B^*_{lm}; \quad
\tilde S_l^{A,B}\nfw = \sum_{l'} M_{ll'} S_l^{A,B}\nfw; \quad
M_{ll'} = {1 \over 2l + 1}\sum_{l'l''} I^2_{ll'l''} |w_{l''}|^2;
\quad
\label{eq:alm_est}\\
&& \hat S_l^{A,B}\nfw = \sum_{l'} [M^{-1}]_{ll'} \tilde S_{l'}^{A,B}\nfw; \quad
\la \delta \hat S_l^{A,B}\nfw \delta \hat S_{l'}^{A,B}\nfw \ra =
\sum_{LL'}M^{-1}_{lL} \la \delta \tilde S_{L}^{A,B}\nfw \delta \tilde S_{L'}^{A.B}\nfw \ra M^{-1}_{L'l'};
\quad \langle \hat S_l^{A,B}\nfw \rangle = S_l^{A,B}\nfw;  \\
&& \delta S_l^{A,B}\nfw =  \la \hat S_{l'}^{A,B}\nfw \ra -
S_l^{A,B}\nfw; \quad\quad\quad\quad \left \{ A,B \right \} \in \left
\{\myf,\myf^2, (\nabla \myf\cdot \nabla \myf), \nabla^2 \myf
\right\}. \label{eq:auto_cov} \een The final expressions will be
independent of the azimuthal quantum number $m$ due to our
assumption of isotropy. In the above derivation we have used the
orthogonality properties of the $3j$ symbols.

It is interesting to note that the {\em convolved} power spectrum
estimated from the masked sky is a linear combination of all-sky
spectra and depends only on the power spectrum of the mask used. The
linear transform is encoded in the mode-mode coupling matrix
$M_{ll'}$ which is constructed from the knowledge of the power
spectrum of the mask. In certain situations where the sky coverage
is very restricted the direct inversion of the mode-mixing matrix
$M$ may not be possible sue to its singularity and binning may be
essential. Based on these results it is possible to define an
unbiased estimator that we denote by $\hat S_l^{A,B}\nfw$. The
noise, due to its Gaussian nature, does not contribute in these
estimators which remains unbiased. However, the presence of noise is
felt in the increase in the scatter or covariance of these
estimators (which can be computed analytically):
\be
[\hat V_k^{(2)}]_l =  \sum_{l'} [M^{-1}]_{ll'} [\tilde V_k^{(2)}]_l; \quad \quad
\la \delta \hat V_k^{(2)} \delta \hat V_{k'}^{(2)} \ra =
\sum_{LL'}M^{-1}_{lL} \la \delta [\tilde V_{k}^{(2)}]_l \delta [\tilde V_{k'}^{(2)}]_{l'} \ra M^{-1}_{L'l'}.
\ee
\n
The symbol $S_l^{A,A}\nfw$ denotes the power spectrum of the field $A(\oh)$; $A(\oh)$ is a generic field
that are used for the construction of generalized skew-spectra.  The derivation of the
covariance depends on a Gaussian approximation i.e. we ignore higher-order non-Gaussianity in the fields.
$\myC_l$ is the ordinary CMB power spectra it also includes the effect of instrumental noise and beam $\myC_l\nfw = \myC^S_lb_l^2\nfw + n_l$.
For a survey with homogeneous noise, we can write $\myC_l^N = \Omega_p \sigma_N^2$ where
$\Omega_p$ is the pixel area and $\sigma_N$ is the noise r.m.s. In a regime when noise contributions dominate, the
MFs can be approximated by a Gaussian approximation. The resulting expressions are listed below:
\ben
&& \la \delta S_l^{A,B}\nfw \delta S_{l'}^{A,B}\nfw \ra = {1 \over 2l+1}
\left [ S_l^{A,A}\nfw \; S_l^{B,B}\nfw + [S_l^{A,B}\nfw]^2 \right ];\label{eq:err1} \\
&& \la \delta S_l^{A_1,B_1}\nfw \delta S_{l'}^{A_2,B_2}\nfw \ra =  {1 \over 2l+1}  \left [ S_l^{A_1,A_2}\nfw\;S_l^{B_1,B_2}\nfw +S_l^{A_1,B_2}\nfw\;S_l^{A_2,B_1}\nfw \right ]; \quad\\
&& \quad\quad\quad \left \{ A_1,B_1,A_2,B_2 \right \} \in \left
\{\myf,\myf^2, (\nabla \myf\cdot \nabla \myf), \nabla^2 \myf
\right\}. \label{eq:err2} \een In the next section we will provide
detailed explicit expressions for various choices of estimators.

The two-to-one estimators are from a family of non-Gaussian
estimators. The three-to-one estimator probes the four-point
correlation function or equivalently the (angular) trispectrum
$T^{l_1l_2}_{l_3l_4}(L)$ \citep{Cooray8}. These spectra have been
used to probe primordial non-Gaussianity beyond lowest order i.e. to
separate contributions from $\tau_{\rm NL}$ and $g_{\rm NL}$. The
two-to-two spectrum or the power-spectrum of the squared CMB maps
was found to be useful in the context of studies of weak lensing
induced CMB non-Gaussianity \citep{CooKes03}. The next order
corrections to the generalized skew-spectra too have been
constructed using such an approach \citep{MSC10} in the context of
studies of MF. The Pseudo-${\cal C}_l$ formalism discussed above is
useful for constructing a  quadruplets of trispectra. The error
estimates for these higher order contributions too can be computed
using a formalism we outline next.
\section{Explicit Expressions for Covariances}
As we have already stressed, the estimators we have introduced for
the skew-spectra are correlated and do not carry independent
information. Their correlation structure depends on the experimental
beam, noise and sky-coverage. Just as with non-Gaussianiaty, partial
sky coverage also introduces mode-mode coupling. However using the
mode-mixing matrix defined in Eq.(\ref{eq:alm_est}) it is possible
to deconvolve the convolved skew-spectra $\tilde S_l$. In this
section, we list the expressions for the co-variance of various
estimators for skew-spectra. The variances or the scatter of the
skew-spectra defined in Eq.(\ref{eq:auto_cov}) take the following
forms:
\ben
&& \la [\delta S_l^{\Theta^2,\Theta}\nfw]^2 \ra \equiv \la [S_l^{\Theta^2,\Theta}\nfw]^2\ra - \la S_l^{\Theta^2,\Theta}\nfw \ra^2 =
{f_{\rm sky}^{-1}\over 2l+1}\left [ S_l^{\Theta^2,\Theta^2}\nfw\; S_l^{\Theta,\Theta}\nfw + [S_l^{\Theta^2,\Theta}\nfw]^2 \right ], \label{eq:var1}\\
&& \la [\delta S_l^{\Theta^2,\nabla^2\Theta}\nfw]^2 \ra
 \equiv \la [S_l^{\Theta^2,\nabla^2\Theta}\nfw]^2\ra - \la S_l^{\Theta^2,\nabla^2\Theta}\nfw \ra^2 ={f_{\rm sky}^{-1}\over 2l+1}\left [
 S_l^{\Theta^2,\Theta^2}\nfw \; S_l^{\nabla^2\Theta,\nabla^2\Theta}\nfw +  [S_l^{\Theta^2,\nabla^2\Theta}\nfw]^2 \right ], \label{eq:var2} \\
&& \la [\delta S_l^{\nabla^2\Theta,\nabla\Theta\cdot\nabla\Theta}\nfw]^2 \ra
 \equiv \la [S_l^{\nabla^2\Theta,\nabla\Theta\cdot\nabla\Theta}\nfw]^2\ra - \la S_l^{\nabla^2\Theta,\nabla\Theta\cdot\nabla\Theta}\nfw \ra^2 \nn \\
&& \quad\quad\quad\quad\quad\quad = {f_{\rm sky}^{-1}\over
2l+1}\left [ S_l^{\nabla^2\Theta,\nabla^2\Theta}\nfw
\;S_l^{\nabla\Theta\cdot\nabla\Theta,\nabla\Theta\cdot\nabla\Theta
}\nfw+[S_l^{\nabla^2\Theta,\nabla\Theta\cdot\nabla\Theta}\nfw]^2
\right ]. \label{eq:var3} \een Here $f_{\rm sky}$ is the fraction of
the sky covered. The expressions for the cross-covariances can also
be computed using the similar technique. We express the harmonics of
the composite fields such as $[\Theta^2]_{lm}$,
$[\nabla\Theta\cdot\nabla\Theta]_{lm}$ in terms of harmonics of the
$[\Theta]_{lm}$ field using the overlap integral. The  final
equations are derived using Wick's theorem to simplify the resulting
expressions. \ben && \la \delta S_l^{\Theta^2,\Theta}\nfw\; \delta
S_l^{\nabla \Theta \cdot \nabla \Theta,\Delta^2\Theta}\nfw \ra =
{f_{\rm sky}^{-1}\over 2l+1}\left [ [S_l^{\Theta^2,\nabla \Theta
\cdot \nabla \Theta}\nfw][S_l^{\Theta,\Delta^2\Theta}\nfw]
+ [S_l^{\Theta,\nabla\Theta \cdot \nabla\Theta}\nfw][S_l^{\Theta^2,\Delta^2\Theta}\nfw] \right ], \label{eq:cross1}\\
&& \la \delta S_l^{\Theta^2,\Theta}\nfw\; \delta S_l
^{\Theta^2,\Delta^2\Theta}\nfw \ra = {f_{\rm sky}^{-1}\over
2l+1}\left [
S_l^{\Theta^2,\Theta^2}\nfw\;S_l^{\Theta,\Delta^2\Theta}\nfw
+ S_l^{\Theta^2,\Theta}\nfw\;S_l^{\Theta^2,\Delta^2\Theta}\nfw \right ], \label{eq:cross2}\\
&& \la  \delta S_l^{\nabla\Theta
\cdot\nabla\Theta,\Delta^2\Theta}\nfw \; \delta S_l
^{\Theta^2,\Delta^2\Theta}\nfw \ra = {f_{\rm sky}^{-1}\over
2l+1}\left [ S_l^{\Theta^2,\nabla \Theta \cdot
\nabla\Theta}\nfw\;S_l^{\Delta^2\Theta,\Delta^2\Theta}\nfw +
S_l^{\Theta^2,\Delta^2\Theta}\nfw
S_l^{\nabla\Theta\cdot\nabla\Theta,\Delta^2\Theta}\nfw \right ].
\label{eq:cross3} \een
The final expressions that we derive are applicable to near all-sky
surveys. When a small portion of the sky is covered a sky patch
version of our calculations can be performed using two dimensional
Fourier analysis instead of the spherical harmonic analysis that we
use here.  Some of the terms appearing in these expressions can be
expressed in terms of the bispectrum. If we assume that the
instrumental noise is Gaussian then there is no contribution from
noise in these expressions.
\ben
&& S_l^{\la \Theta^2, \nabla^2\Theta \ra}(\fw) =-\sum_{l_i=2}^{l_{\rm max}} l(l+1) \; B_{ll_1l_2}J_{ll_1l_2}b_{l}b_{l_1}b_{l_2}, \\
&& S_l^{\la \Theta, \nabla\Theta\cdot\nabla\Theta \ra}(\fw)=
\sum_{l_i=2}^{l_{\rm max}} B_{ll_1l_2} J_{ll_1l_2}[l(l+1)+l_1(l_1+1)-l_2(l_2+1)]b_{l}b_{l_1}b_{l_2},\\
&& S_l^{\la \nabla\Theta\cdot\nabla\Theta, \Delta^2\Theta \ra}(\fw)
= -\sum_{l_i=2}^{l_{\rm max}} B_{ll_1l_2} J_{ll_1l_2} \{
l(l+1)[l(l+1)+l_1(l_1+1)-l_2(l_2+1)] \} b_{l}b_{l_1}b_{l_2}. \een
Notice that these expressions are generic, in that they are derived
without any specific assumption about the shape of the bispectrum.
The rest of the terms can be expressed in terms of the power
spectrum alone. As is common practice in the literature these
results ignore all higher order correlation beyond the bispectrum.
\ben && S_l^{\la \Theta^2, \nabla\Theta\cdot\nabla\Theta \ra}(\fw) =
{1 \over 2l+1}\sum_{l_i=2}^{l_{max}}\left ( \begin{array}{ c c c }
     l_1 & l_2 & l \\
     0 & 0 & 0
  \end{array} \right)^2 I^2_{l_1l_2l} [l(l+1)+l_1(l_1+1)-l_2(l_2+1)]
(\myC_{l_1} b_{l_1}^2+n_{l_1})(\myC_{l_2} b_{l_2}^2+n_{l_2}), \\
&& S_l^{\la \Theta^2,\Theta^2 \ra} ={2 \over 2l +1 } \sum_{l_i=2}^{l_{max}}\left ( \begin{array}{ c c c }
     l_1 & l_2 & l \\
     0 & 0 & 0
  \end{array} \right)^2 I^2_{l_1l_2l}(\myC_{l_1}b^2_{l_1}+n_{l_1})(\myC_{l_2}b^2_{l_2} + n_{l_2}),\\
&& S_l^{\la \nabla\cdot\nabla,\nabla\cdot\nabla \ra} ={2 \over 2l+1} \sum_{l_i=2}^{l_{max}}\left ( \begin{array}{ c c c }
     l_1 & l_2 & l \\
     0 & 0 & 0
  \end{array} \right)^2 I^2_{l_1l_2l}[l(l+1)+l_1(l_1+1)-l_2(l_2+1)]^2(\myC_{l_1}b^2_{l_1} +n_{l_1})(\myC_{l_2}b^2_{l_2} + n_{l_2}).
\een
The remaining terms are scaled input power-spectra:
\be
S_l^{\la \Theta, \nabla^2\Theta \ra} = -l(l+1)(\myC_l b_l^2 +n_l); \quad
S_l^{\la \nabla^2\Theta, \nabla^2\Theta \ra} = l^2(l+1)^2(\myC_l b_l^2 +n_l); \quad
S_l^{\Theta,\Theta} =(\myC_l b_l^2 +n_l).
\ee
The derivation of these results follow the same general principle that is outlined
in \textsection\ref{sec:estim}.
These expressions are used to compute the cross-correlation coefficient among
various spectra which are defined below:
\ben
&& r^{ij}_l(\fw) = \la \delta S_l^{(i)}(\fw) \delta S_l^{(j)}(\fw)\ra /
\sqrt{ \la [\delta S_l^{(i)}(\fw)]^2\ra} \sqrt{ \la [\delta S_l^{(j)}(\fw)]^2 \ra}  ;
\quad\quad i,j \in \{0,1,2\}.
\label{eq:cross_def}
\een
As before, throughout we have ignored the mode-mode coupling. The coefficients of cross-correlation  $r_{ij}$ are
independent of the sky-coverage $f_{\rm sky}$. The signal to noise for individual modes for a given spectrum
on the other hand can be expressed as:
\ben
&& [{S/N}]^{(i)}_l(\fw) =  \sqrt{\la [S_l^{(i)}(\fw)]^2\ra /\la [\delta S_l^{(i)}(\fw)]^2\ra }\quad\quad i \in \{0,1,2\}.
\label{eq:s2n_def}
\een
The cumulative  $[S/N]=\sum_{l=2}^{l_{\rm max}} [{S/N}]^{(i)}_l$ is tabulated for individual experiments
in Table-\ref{table:signal_noise1} for Planck and EPIC.

\begin{figure}
\begin{center}
{\epsfxsize=15 cm \epsfysize=4.8 cm {\epsfbox[37 531 589 714]{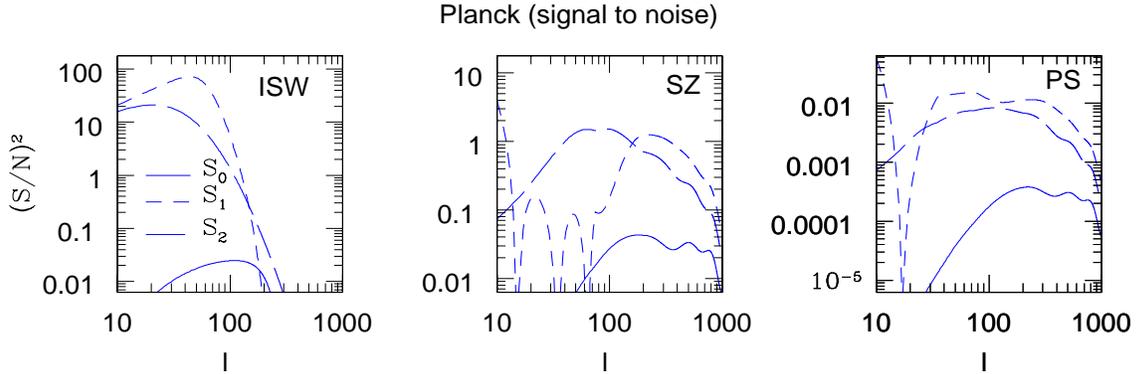}}}
\end{center}
\caption{The mode-by-mode signal to noise ratio is depicted for the
three skew-spectra, $S^{(0)}_l\nfw$ (left-panel), $S^{(1)}_l\nfw$
(middle-panel) and $S^{(2)}_l\nfw$ (right-panel), as a function of
the harmonics $l$. Two different experimental set-ups are considered
ongoing experiment Planck (solid curves) and EPIC (dashed-curves).
The parameters defining these experiments are summarized in Table
\ref{tab:exp}. We have assumed an all-sky coverage $f_{\rm sky}=1$.
The mixed bispectrum that the skew-spectra sample correspond to
cross-correlation of lensing and thermal SZ effect. All other source
of secondary and primordial non-Gaussianity has been ignored. We use
Eq.(\ref{eq:err1}) in conjunction with Eq.(\ref{eq:err2}) in our
computation of the scatter.\label{fig:s2n_planck}}
\end{figure}

\begin{figure}
\begin{center}
{\epsfxsize=15 cm \epsfysize=4.8 cm {\epsfbox[37 531 589 714]{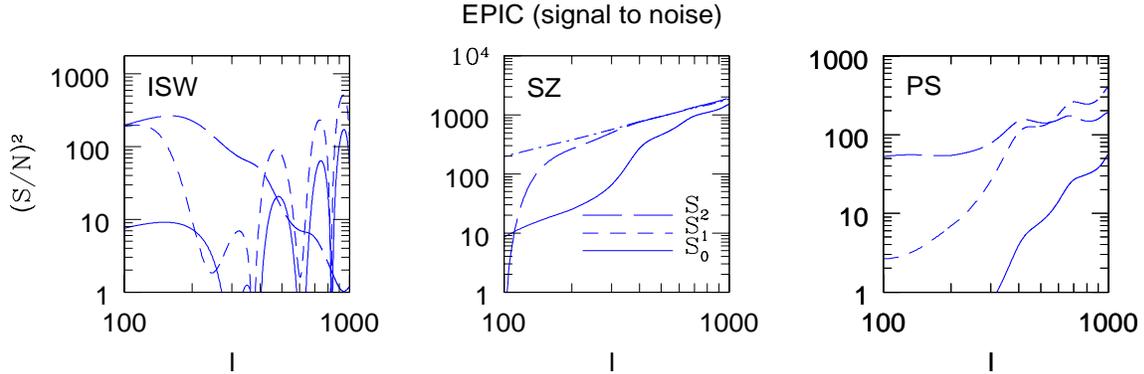}}}
\end{center}
\caption{The signal to noise ratio for the three skew-spectra,
$S^{(0)}_l\nfw$ (left-panel), $S^{(1)}_l\nfw$ (middle-panel) and
$S^{(2)}_l\nfw$ (right-panel), are depicted as a function of the
harmonics $l$. Two different experimental set-ups are considered
ongoing experiment Planck (solid curves) and EPIC (dashed-curves).
The parameters defining these experiments are summarized in Table
\ref{tab:exp}. We have assumed an all-sky coverage $f_{\rm sky}=1$.
The mixed bispectrum that the skew-spectra sample correspond to
cross-correlation of lensing and thermal SZ effect. All other source
of secondary and primordial non-Gaussianity have been
ignored.\label{fig:s2n_epic}}
\end{figure}

We have also computed the signal to noise ratio for individual modes
using these expressions for various skew spectra. These results are
plotted in Figure \ref{fig:s2n_epic} for EPIC as well as for  Planck
in Figure (\ref{fig:s2n_planck}). The cumulative signal-to-noise as
expected is higher for EPIC due to higher sensitivity. The signal to
noise for ISW decreases sharply at higher $l$ and peak at lower $l$
on the other hand the signal to noise for SZ and unresolved point
sources peak at a much higher angular frequency. Among the three
skew-spectra we have considered, the skew spectra
$S^{(1)}_l(\theta_b)$ was found to have higher signal to noise
compared to $S^{(0)}_l(\theta_b)$ and $S^{(2)}_l(\theta_b)$. While
the lowest order skew-spectra $S^{(0)}_l(\fw)$ is dominated mostly
by cosmic variance the other skew-spectra, $S^{(2)}_l(\theta_b)$ is
maximally affected by the instrumental noise. The information
content is not independent for the different skew-spectra; their
cross-correlation coefficient provides a succinct measure of this
lack of dependence.

To correct for the effect of a mask and the noise we will follow the
Pseudo-${\cal C}_{\ell}$ (PCL) method devised by \cite{Hiv} for
power spectrum analysis and later developed by \cite{MuVaCoHe11} for
analyzing the skew spectra and the kurt-spectrum \citep{Mu11c}.and
the cross-correlation coefficient provides a valuable indicator of
their independence.

The signal to noise of estimates of one-point generalised skewness
parameters $S^{(i)}=\sum_{l=2}^{l_{\rm max}} (2l+1)S^{(i)}_l$ is
given by $\left [\sum_{l=2}^{l_{\rm max}}{(2l+1)^2\la[\delta
S^{(i)}/S^{(i)}]^2\ra} \right ]^{-1/2}$. The corresponding numbers
for Planck and EPIC are presented in Table
\ref{table:signal_noise2}.

%
It is possible to introduce a filter function $w_{l_1l_2}$ in the
definition of the skew-spectra. Choices include, sharp cutoff in the
$l$ space to avoid the affect of noise at high $l$, to optimal
filters that maximizes the signal to noise for a given resolution
$l_{\rm max}$. Clearly such option will invariably improve the
statistical significance. The filter functions can be of further
interest if the bispectrum is more pronounced for certain triangular
configurations. Another potentially useful application of a filter
function is to filter out a specific configuration. However in such
case the skew-spectra will not have any direct relation with the
topological properties of the original map.
\section{Conclusion and Discussion}
\label{sec:concl}
Non-Gaussianity is in itself a poorly defined concept, in that there
is no unique approach that can be adopted to describe or parametrize
an arbirary form of non-Gaussianity in a complete manner. In order
to quantify non-Gaussianity as fully as possible it is therefore
essential to deploy a battery of complementary approaches  each of
which exploits different statistical characteristics. Each such
technique will have a unique response to real world issues such as
the sky-coverage (observational mask), beam and instrumental noise.
Any robust detection therefore will have to involve a simultaneous
cross-validation of results obtained from independent methods. The
most common characterizations of non-Gaussianity involve studying
the bispectrum, which represents the lowest-order departure from
Gaussianity; higher order non-Gaussianity can be studied using its
higher order analogues i.e. the multi-spectra.

By contrast the topological estimators (MFs) that we have studied
here carry information to all orders, though in a collapsed
(one-point) form. Analytical results for MFs for a Gaussian field
are well understood, and form the basis of non-Gaussianity studies
\citep{Tomita86}. There have been several previous studies on
extraction of the MFs from the CMB data that rely either on
simplification of radiative transfer using the Sachs-Wolfe limit
\citep{HKM06} or using a perturbative approach based on a series
expansion of the MFs that can be studied order by order
\citep{Mat03,mat94}. The MFs have also been studied using elaborate
computer-intensive non-Gaussian simulations
\citep{Komatsu03,Spergel07}. Most of these studies were done using a
specific model of  non-Gaussianity, namely the {\em local} model of
primordial non-Gaussianity which is parametrized by the well known
parameter $f_{\rm NL}$.

The main motivation behind the present study has been to to extend
such methods to secondary non-Gaussianities which have not been
studied before in the context of morphological statistics
analytically. The increase in sensitivity of CMB experiments and
near all-sky coverage along with wide frequency range means the
study of non-Gaussianities will be feasible in the very near future.
Moreover, in the currently favoured adiabatic CDM models it is
expected that the contribution from primordial non-Gaussianity is
negligible and the main contribution to non-Gaussianities comes from
secondaries. The secondary non-Gaussianity signal are associated
with large scale structure contributions and through various mode
coupling effects such as gravitational lensing
\citep{GoldbergSpergel99a,GoldbergSpergel99b, CoorayHu}. Our primary
aim in this work has been to study how well we can probe the
secondary signals from mode coupling using morphological
descriptors.

One of the main difficulties faced by one-point estimators
$S^{(i)}(\fw)$, which also affects the MF-based estimators
$V_k(\nu)$, is their inability to differentiate among various
sources of non-Gaussianity. The triplets of skew spectra
$S^{(i)}_l(\fw)$ that we have introduced can be used to separate out
contributions from various secondaries as well as to probe and
constrain any foreground residuals left from the component
separation step of the data analysis chain. Generalizing
\cite{MuHe10} we have introduce a set of triplets of skew-spectra
which can be extracted from any realistic data. These skew-spectra
do not compress the available information from a bispectrum to a
single number, and their shape can help to distinguish among various
sources of non-Gaussianity. Exploiting the perturbative expansion of
the MFs, we showed that at the leading order of non-Gaussianity the
MFs are completely specified by the knowledge of the bispectrum. Our
results are most naturally defined in the harmonic domain.
Comparison of MFs extracted using harmonic approach can be
cross-compared with more traditional approach in the real space as
an useful consistency check.

The methods based on the skew-spectra that we have presented are
simple to implement once the derivative fields $[\nabla
\Theta\cdot\nabla\Theta]$ or $[\nabla^2\Theta]$ are constructed.  We
have shown that this can be implemented in a model-independent way.
Our method is based on a Pseudo-${\cal C}_l$ approach \citep{Hiv}
and can handle arbitrary sky coverage and inhomogeneous noise
distributions. The Pseudo-${\cal C}_{\ell}$ approach is well
understood in the context of power spectrum studies, and its
variance or scatter can be computed analytically. We provide generic
analytical results for the computation of scatter around individual
estimates. We also provide detailed predictions on how they are
cross-correlated. In our method, it is possible indeed to go beyond
the lowest level in non-Gaussianity to include the contribution from
trispectrum. The main contributions in frequency-cleaned CMB maps
will be from lensing of the CMB, though it is expected that such
corrections will be sub dominant at least in the context of CMB data
analysis.
%
%
%
%
\begin{table*}
\begin{center}
\begin{tabular}{|c |c|c| c| c}
\hline
(Planck,EPIC) & SZ  & ISW & PS &\\
\hline
$\left ({S/N} \right )$ & $(5.0,1137.4)$ & $(1.0, 216.0)$ & $(0.5,209.0)$ \\
\hline
$\left ({S/N} \right )$ & $(24.0,1354.9)$ & $(62.2, 420.3)$ & $(4.3,552.0)$\\
\hline
$\left ({S/N} \right )$ & $(19.7,1328.8)$ & $(31.8,246.5)$ & $(1.7, 421.0)$ \\
\hline
\end{tabular}
\caption{The cumulative signal to noise (S/N) for Planck and EPIC surveys, are shown for
the three one-point skew-spectra. Parameters used to compute the skew-spectra and the associated scatter for the two
different experiments, ongoing Planck \citep{Planck08} and EPIC \citep{Bau09}.\label{table:signal_noise1}}
\end{center}
\end{table*}
%
%
\begin{table*}
\begin{center}
\begin{tabular}{|c |c|c| c| c}
\hline
(Planck,EPIC)  & SZ  & ISW & PS &\\
\hline
$\left ({S/N} \right )$ & $(3.8,503.2)$ & $(0.3,39.3)$ & $(0.4,53.4)$ \\
\hline
$\left ({S/N} \right )$ & $(5.8,1299.0)$ & $(6.4\times 10^{-2},70.9)$ & $(0.6,529.1)$ \\
\hline
$\left ({S/N} \right )$ & $(1.2,625.8)$ & $(1.4\times 10^{-2},21.1)$ & $(0.1,178.2)$ \\
\hline
\end{tabular}
\caption{The cumulative signal to noise (S/N) for Planck and EPIC
surveys for the one-point cumulants $S^{(i)}(\oh)$, defined in
Eq.(\ref{eq:s2n_def}), are shown for each skew-spectra. Parameters
used to compute the skew-spectra and the associated scatter for the
two different experiments, ongoing Planck \citep{Planck08} and EPIC
\citep{Bau09}. We have assumed $f_{\rm sky}=1$ in our calculations.
\label{table:signal_noise2}}
\end{center}
\end{table*}
%
%
\begin{figure}
\begin{center}
{\epsfxsize=4.8 cm \epsfysize=5.5 cm {\epsfbox[37 439 294 714]{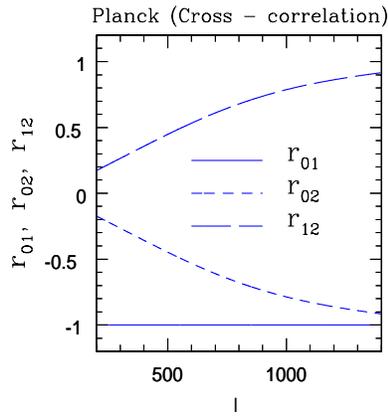}}}
\end{center}
\caption{Same as the previous Figure but for Planck noise. The
expression for scatter Eq.(\ref{eq:var1})-Eq.(\ref{eq:var3})and
cross-correlation Eq.(\ref{eq:cross1})-Eq.(\ref{eq:cross3}) has two
contributions. In each of these expressions there are terms which
depend on the bispectrum and there are terms which can constructed
from power spectrum alone. For Planck noise we found that the
expressions for the scatter as well as the cross-correlation are
entirely dominated by the terms which depend only on the power
spectrum thus making the coefficient $r_{ij}$ independent of the
type of underlying bispectrum.} \label{fig:cross_planck}
\end{figure}
\begin{figure}
\begin{center}
{\epsfxsize=15 cm \epsfysize=4.8 cm {\epsfbox[37 531 589 714]{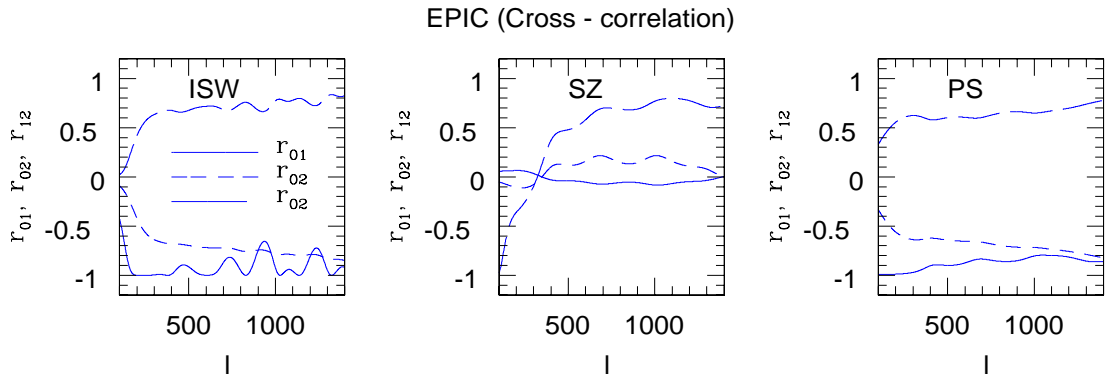}}}
\end{center}
\caption{The information content of the skew-spectra are not
completely independent. The level of cross-correlation among various
estimator is encoded in the coefficient of cross-correlations
$r_{ij}$ defined in Eq.(\ref{eq:cross_def}). The cross-correlation
coefficient $r_{01}$, $r_{02}$ and $r_{12}$ are plotted as a
function of harmonics $l$. The noise level correspond to that of
EPIC (see Table \ref{tab:exp}). The correlation structure refelects
the underlying spectra $\beta_l$ as well as the level of noise. The
cross-correlation is similar for SZ and PS.} \label{fig:cross_epic}
\end{figure}

We conclude by pointing out that the MFs do not probe the full
bispectrum, but involve only weighted sums of modes and are thus
equivalent to the three {\it generalised} skewness parameters we
have used. We have also defined three generalized skew-spectra
associated with each of these skewness parameters. In this sense,
the study of these skew-spectra can replace the study of MFs. The
skew-spectra we have introduced can all be probed for arbitrary mask
and noise. Ubiased estimators can also be constructed which can work
in the presence of partial sky coverage and inhomogeneous noise.
Their variance can also be computed {\em analytically}; thereby
avoiding the use of non-Gaussian Monte-Carlo simulations completely.
Finally, the MFs can be constructed from the knowledge of
generalized skew-spectra and can be compared with the results from
real space analysis. The triplets of generalized skew-spectra can be
used to separate individual components of NGs using their shape
information. From our analytical results of cross-correlation, we
find that in the absence of noise, e.g. experiments such as EPIC,
the skew-spectra are highly correlated, more so for higher $l$
values. The correlation coefficients are typically in the range $r =
0.5-1$ for a Planck type experiment. The cumulative signal-to-noise
ratio, in a Planck type experiment, for bispectrum corresponding to
the ISW and SZ and lensing cross-correlation reaches ${\cal O}(10)$.
An improvement of about two orders of magnitude can be expected with
experiments such as EPIC.

Throughout we have ignored the presence of primordial
non-Gaussianity which is expected to be subdominant. Nevertheless,
it can be incorporated. Individual skew-spectra from different
underlying bispectrum can essentially be combined to construct the
total skew-spectra which means that our results can straightfowardly
be generalised to incorporate specific models of primordial
non-Gaussianity.

The generic results derived here are also applicable to other areas
in cosmology and have indeed been explored recently. Examples
include the analysis of galactic redshift surveys \citep{PratMun12},
weak lensing surveys \citep{MuWaSmCo12} and the frequency cleaned SZ
maps \citep{MuSmJoCo12}. The results presented here can be extended
beyond the analysis of temperature maps, e.g. to analyze
polarisation maps, by extending the spin-$0$ calculations to
spin-$2$. Such results can furnish useful probes for tje
characterization of morphology of reionization in three dimensions.

A few comments are in order about the comparison of our estimators
with the so-called optimal estimators. The motivation to construct
an optimal estimator is to improve the signal-to-noise of detection
which is important in case of weak signals such as the primordial
non-Gaussianity. The main motivation in this paper has been to
reconstruct the topological properties of the CMB map going beyond
Gaussianity, in the harmonic domain; in particular due to the
contributions from secondary lensing cross-correlation which will be
detected with high signal to noise with the proposed CMB surveys
such as EPIC.

In addition to the primary and secondary non-Gaussianity, cosmic
defects such as textures or cosmic strings \citep{ABR99,Cr07,RS10}
also leave non-Gaussian footprints in CMB maps which can be detected
by the change in topological nature of the maps. The estimators we
have presented here may have relevance in such investigations. A
detailed study will be presented elsewhere.

At the level of the bispectrum the effect of lensing can only be
studied through its cross-correlation with other secondaries.
However weak lensing is also independently responsible for a next
order correction to MFs through its effect on the trispectrum; the
signal-to-noise is expected to be low.

The signal-to-noise of the skew-spectra for secondary-lensing
cross-correlation bispectrum is comparable to that of the
skew-spectra of frequency-cleaned SZ maps \citep{MuSmJoCo12}.
However the secondary skew-spectra are much higher compared to
skew-spectra associated with primary skew-spectra unless we assume a
rather high value for the $f_{\rm NL}$.
\section{Acknowledgements}
\label{acknow} DM and PC acknowledges support from STFC standard
grant ST/G002231/1 at the School of Physics and Astronomy at Cardiff
University where this work was completed. DM would like to thank
Joseph Smidt, Geraint Pratten, Asantha Cooray, Shahab Joudaki and
Erminia Calabrese for very useful discussions.
\bibliography{paper.bbl}
\appendix
\section{Spherical Harmonics}
The completeness relationship for the spherical harmonics $Y_{lm}(\oh)$ is given by:
\ben
\sum_{lm} Y_{lm}(\oh)Y^*_{lm}(\oh') = \delta_{2\rm D}(\oh-\oh').
\label{complete_spherical}
\een
The orthogonality relationship is as follows:
\ben
\int d\oh \; Y_{lm}(\oh) Y^*_{l'm'}(\oh) = \delta^{\rm K}_{ll'}\delta^{\rm K}_{mm'}.
\label{ortho_spherical}
\een
Here $\delta_{2\rm D}$ and  $\delta^{\rm K}$ represents the Dirac's 2-dimensional delta function and Kroneker's Delta function resepctively.
\section{3j Symbols}
The following properties of $3j$ symbols were used to simplify various expressions.
\ben
\sum_{l_3m_3} (2l_3+1) \left ( \begin{array}{ c c c }
     l_1 & l_2 & l_3 \\
     m_1 & m_2 & m_3
  \end{array} \right )
\left ( \begin{array}{ c c c }
     l_1 & l_2 & l \\
     m_1' & m_2' & m
  \end{array} \right ) = \delta^K_{m_1m_1'} \delta^K_{m_2m_2'};
\een
\ben
\sum_{m_1m_2} \left ( \begin{array}{ c c c }
     l_1 & l_2 & l_3 \\
     m_1 & m_2 & m_3
  \end{array} \right)
\left ( \begin{array}{ c c c }
     l_1 & l_2 & l_3' \\
     m_1 & m_2 & m_3'
  \end{array} \right) = {\mathcal \delta^K_{l_3l_3'} \delta^K_{m_3m_3'} \over 2l_3 + 1};
\label{3j_ortho1}
\een
\ben
(-1)^m\left ( \begin{array}{ c c c }
     l & l & l' \\
     m & -m & 0
  \end{array} \right ) = {(-1)^l \over \sqrt{(2l+1)}} \delta^K_{l'0};
\label{eq:3j}
\een
\ben
\int d\oh Y_{lm}(\oh)Y_{l'm'}(\oh)Y_{LM}(\oh) = \sqrt{(2l+1)(2l+1)(2L+1)\over 4\pi}\left ( \begin{array}{ c c c }
     l & l' & L \\
     m & m' & M
  \end{array} \right)
\left ( \begin{array}{ c c c }
     l & l' & L \\
     0 & 0 & 0
  \end{array} \right).
\label{Gaunt}
\een
\end{document}